\documentclass[pre,amssymb,twocolumn,showpacs,superscriptaddress,floatfix]{revtex4}

\usepackage{graphicx}
\usepackage{bm}
\usepackage{mathptmx}

\begin{document}

\title{How to become a superhero}

\author{P. M. Gleiser}
\affiliation{Centro At\'omico Bariloche and Instituto Balseiro, 8400 S. C. de Bariloche, Argentina\\
 Consejo Nacional de Investigaciones Cient\'{\i}ficas y T\'ecnicas, Argentina.}

\begin{abstract}
We analyze a collaboration network based on the Marvel Universe comic
books. First, we consider the system as a binary network,  where two characters 
are connected if they appear in the same publication. The analysis of degree correlations
reveals that, in contrast to most real social networks, the Marvel Universe presents a
disassortative mixing on the degree. Then, we use a weight  measure to study
the system as a weighted network. This allows us to find and
characterize well defined communities. Through the analysis of the community structure and the
clustering as a function of the degree we show that the network
presents a hierarchical structure. Finally, we comment on possible
mechanisms responsible for the particular motifs observed. 
\end{abstract}

	%Networks and genealogical trees
\date{\today}

\maketitle

\section{Introduction}
In the last years the physics community has devoted a strong effort to
the study and analysis of complex networks~\cite{Dorogovtsev,Albert,Newman,NBW}. These studies
allow for general characterizations such as the small world effect~\cite{Watts} or the scale-free property~\cite{Barabasi} 
which are shared by many systems, including technological, biological and social systems ~\cite{Dorogovtsev,Albert,Newman,NBW,Watts,Barabasi}. 
Among the social networks the so called collaboration networks are of
particular interest given the availability of large databases which allow for extensive statistical analysis, 
and also since the connections between the vertices, which represent individuals, can be precisely defined.
Two well known examples of collaboration networks are the movie actors network, where two actors are connected if 
they appear in the same movie \cite{Watts,Barabasi,Amaral}, and  scientific collaboration networks, where two scientist are 
connected if they are authors in the same  publication~\cite{Newman2,Newman_pre1,Newman_pre2,Barabasi2}.

Perhaps one of the most challenging problems in the characterization
of these systems is the determination  of communities, which can be vaguely defined 
as groups of nodes which are more connected among
themselves than with the rest of the network~\cite{Girvan,NG,Palla,Danon}. 
Through their identification and analysis one can search for fundamental laws in social interactions~\cite{Gleiser,Guimera,Arenas,Boguna,Motter,Noh}. In this article we will work along this line, 
and show that the determination and characterization of the communities allow us to detect mechanisms responsible for the particular 
motifs observed. In particular we will focus on the Marvel Universe (MU)~\cite{MarvelUniverse}
which is  a fictional cosmos created by the Marvel Comics book publishing company. The idea of a common Universe  
allows characters and plots to cross over between publications, and also makes continuous references to events 
that happen in other books. In this Universe real world events
 are mixed with science fiction and fantasy concepts. An interesting question that arises is if
this network, whose nodes correspond to invented entities and 
whose links have been created by a team of writers, resembles in some way real-life social 
networks, or, on the contrary,  looks like a random network. This
issue has been addressed by Alberich, Miro-Julia and Rossell\'o (AMJR)~\cite{Alberich}, which used information from the Marvel Chronology Project database~\cite{Chappell},
to build a bipartite collaboration network. They obtained a network formed by $6486$ characters and $12942$ books,
where two characters are considered linked if they jointly appear in the same
comic book. AMJR found that the MU looks almost as a real social network, since it has 
most of, but not all, the characteristics of real collaboration networks such as movie actors or scientific collaboration networks.
In particular, the  average degree of the MU is much smaller than the theoretical average degree of the 
corresponding random model, thus indicating that Marvel characters collaborate more often 
with the same characters. Also, the clustering coefficient is smaller than what is usual in real
collaboration networks. Finally, the degree distribution presents a power law
with an exponential cutoff,  $P(k) \sim k^{-\tau} 10^{-k/c}$ with an exponent $\tau = 0.7158$. 
Since $\tau$ is much smaller than 2 the average properties of the network are dominated
by the few actors with a large number of collaborators, indicating that some  superheroes 
such as Captain America or Spider Man present much 
more connections than would be expected in a real life collaboration network~\cite{Alberich}. 

\section{Degree Correlations}

\begin{figure}
\includegraphics[width=\linewidth,clip=true]{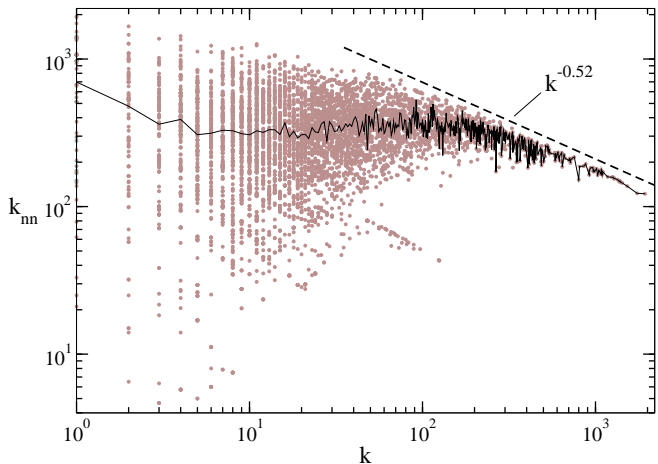}
\caption{\label{figure1} Mean degree $\langle  k_{nn} \rangle$ of the
  neighbors of a vertex as a function of the degree $k$. Circles show the
  different values of $k_{nn}$ for a given $k$. The continuous line indicates 
the average value $\langle  k_{nn} \rangle$.}
\end{figure}

We begin our study by presenting an analysis of the degree correlations which fully reveals the artificial 
nature of the MU. Most real social networks are assortatively mixed by
degree, that is,  vertices with high degree tend to be  connected to vertices with 
high-degree while vertices with low degree tend to be connected to vertices with low degree \cite{Newman4,Newman5}. 
On the other hand, most biological and technological networks present a disassorative mixing by degree, where vertices with high
degree tend to be connected to vertices with low-degree \cite{Newman4}. 
The degree correlations of the network can be analyzed by plotting the mean degree $\langle k_{nn} \rangle$ of the 
neighbors of a vertex as a function of the degree $k$ of that vertex~\cite{Pastor-Satorras}. A positive slope indicates assortative
mixing, while a negative slope signals disassortative mixing on the degree. 
To begin our analysis of the MU we consider the system as a binary network, where two characters are connected if they 
appear in the same publication. In order to compare with their results we use the data compiled by AMJR~\cite{Miro}.  
In Fig. \ref{figure1} we show in  small circles the different values of $k_{nn}$ obtained in the MU for
 a given degree $k$, while the black continuous line shows  the average value $\langle k_{nn} \rangle$.  
For $k<10$  the dispersion in the values of $ k_{nn} $ is too large, 
ranging from a few to more than a thousand, thus not allowing for any characterization of $\langle  k_{nn} \rangle$.  
As $k$ increases the dispersion in the values of  $k_{nn}$ diminishes in a funnel like shape, while $\langle k_{nn}\rangle$
presents fluctuations around a constant value, indicating that no correlations  dominate up to $k \approx 200$. For $k>200$ a 
decreasing  behavior of $ \langle k_{nn} \rangle$ can be clearly observed, and the tail 
seems to follow a power law behavior $\langle k_{nn} \rangle = k^{-\nu}$ with $\nu \approx 0.52$.  
This result shows that, in contrast to what is observed in most
real social networks, the MU network presents a disassortative behavior on the degree. 
Surprisingly, the value of the exponent is similar to the one observed in a real technological
 network: the Internet, where an exponent  $\nu  \approx 0.5$ has been also found~\cite{Pastor-Satorras}.

The origin of a disassortative behavior in the Internet is most probably given by the fact 
that the hubs (nodes with the largest degree) are connectivity providers, and thus have 
a large number of connections to clients that have only a single connection. 
In the MU the small exponent in the  degree distribution observed by AMJR~\cite{Alberich} 
and the disassortative behavior clearly indicate the presence of hubs.
One immediately wonders what is the role that they play in the MU. 
Perhaps the most intuitive idea to answer this question is to make a list that takes into account the degree 
of the characters or to count the number of publications in which they appear. 
A first step in this direction was already taken by AMJR that point
out that Captain America is the superhero with more connections and Spider Man 
is the one that appears in the largest number of comic books~\cite{Alberich}.
Clearly, these classifications help us to establish how popular
  these characters are. However, they do not give information on where to 
establish a cut-off in the ranking list.  Also, when considering the interactions 
between the characters, one is left with the problem on how to deal
  with the large number of connections of these hubs. In the following section 
we tackle these issues. 

\section{The Marvel Universe as a weighted network}

In order to advance a step further in the analysis of the MU, we take into account the 
fact  that some characters appear repeatedly in the  same publications.  
The incorporation of this information allows us  to distinguish
  connections between characters which truly represent a strong social
  tie, such as a connection between two characters that form 
a team, to those  connections that link two characters that perhaps
 have met only once in the whole history of the MU. 
To define the  strength $w_{ij}$ of the ties between characters $i$
and $j$ we use the weight measure proposed by Newman~\cite{Newman_pre2}:
\begin{equation}
w_{ij}= \sum_{k}  \frac{\delta_i^k \delta_j^k}{n_k-1}
\end{equation}
where $\delta_i^k$ is equal to one if character $i$ appears in book $k$ and zero otherwise, 
and $n_k$ is the number of characters in book $k$. 

\begin{figure}
\includegraphics[width=\linewidth,clip=true]{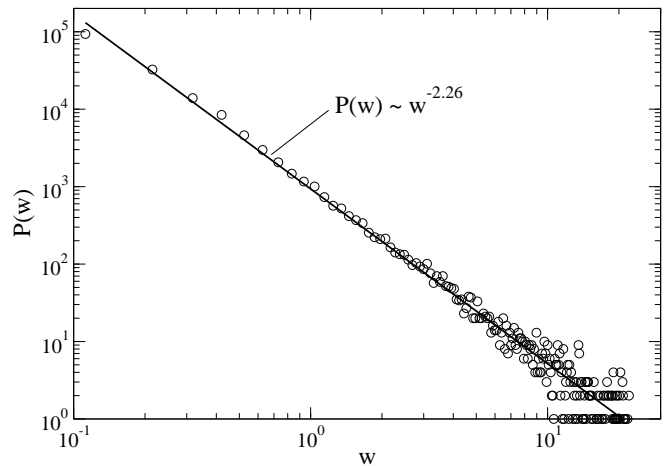}
\caption{\label{figure2} Weight distribution $P(w)$ vs. $w$ (circles). The straight line is a power law fit, $P(w) \sim w^{-2.26}$.}
\end{figure}
In Fig. \ref{figure2} we present the weight distribution $P(w)$ of the
MU network, which can be fitted by a power law, $P(w) \sim w^{-\gamma}$ with $\gamma = 2.26$. 
A power law behavior in weight distributions has also been observed in real
scientific collaboration  networks such as the cond-mat network ($\gamma = 3.7 \pm 0.1$)
and the astro-ph network ($\gamma = 4.0 \pm 0.1$)~\cite{Li}. However, we must point out
that in the MU the distribution extends over more than two decades, while in
the real collaboration networks the distributions reach one decade only. Also, the
value of the exponent in the MU is much smaller. As a consequence, 
a small fraction of the interactions are very strong, while the majority 
interact very weakly. Again we find a result that highlights the leading role 
of a few characters. In this case this is reflected in the fact that they 
interact more frequently than other characters do. 

\begin{figure}
\includegraphics[width=\linewidth,clip=true]{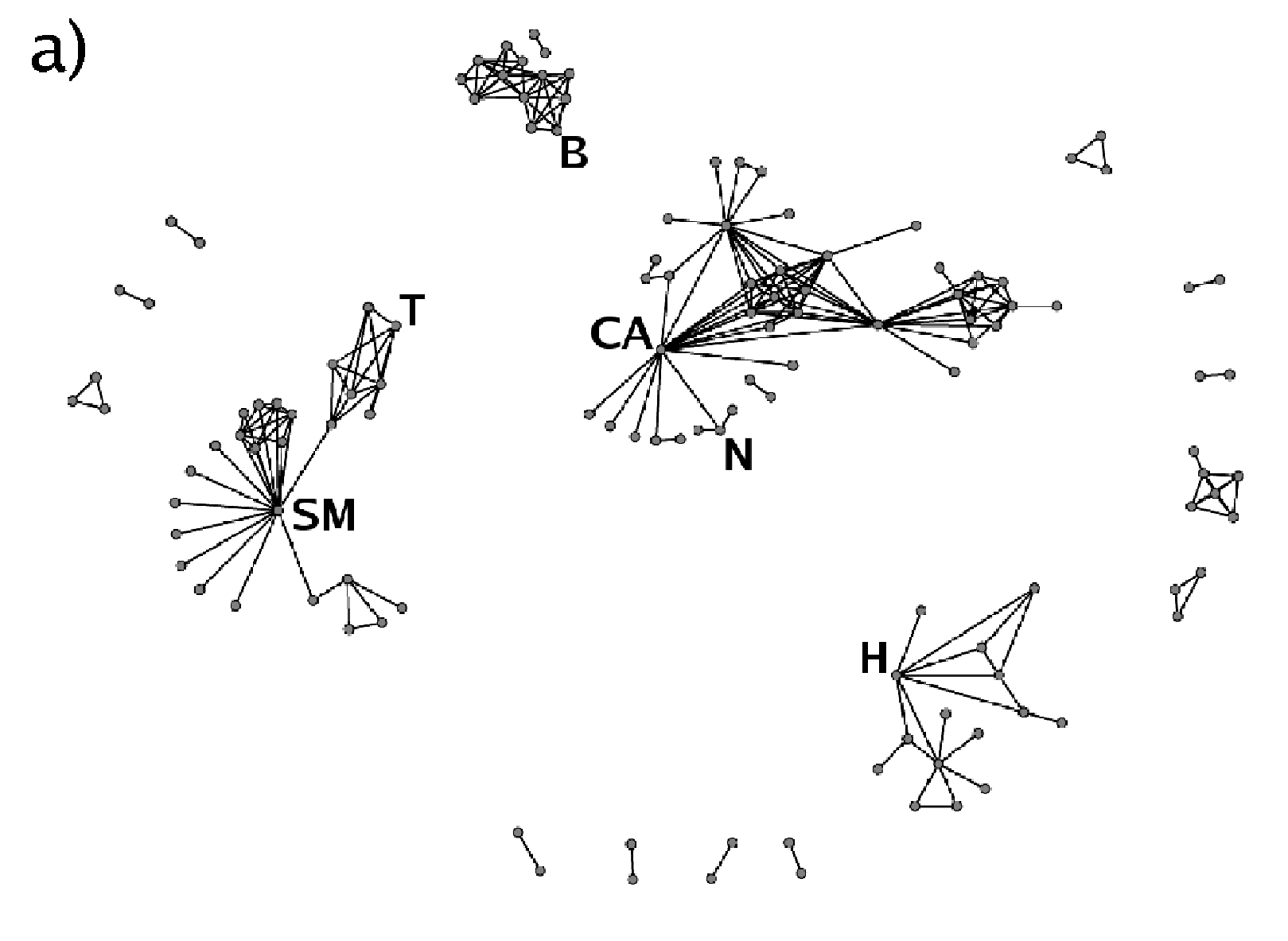}
\includegraphics[width=\linewidth,clip=true]{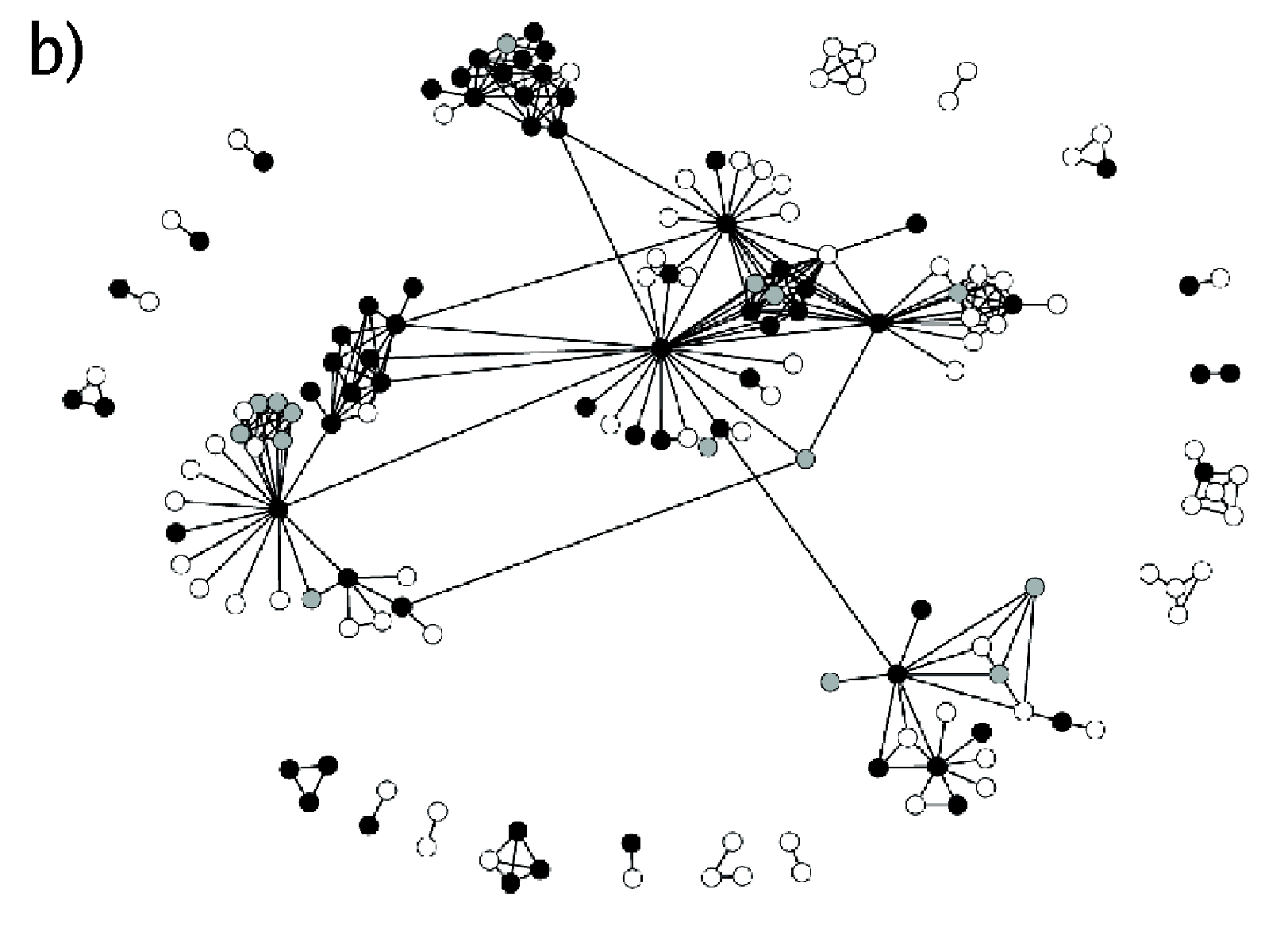}
\caption{\label{figure3}a) Network after the addition of 220
  links. The initials correspond to characters that play an important role connecting 
communities: Spider-Man (SM), Thing (T), Beast (B), Captain America (CA),  Namor (N), Hulk (H). 
b) Network after the addition of 300 links, when a giant component has  emerged.  The black (white) circles 
indicate characters labeled as heroes (villains). The gray circles indicate  other type of characters, such as people, gods or
 nodes with no classification.}
\end{figure}

In order to use the information of the weights to find and characterize the role of these leading characters we  
set a threshold on the weight and consider only those interactions with a value above the threshold. 
When we set the threshold to its highest possible value, in order to leave just the link with the largest 
weight, we find that the connection between Spider Man with his girlfriend (later wife) Mary Jane Watson Parker 
is the strongest in the MU~\cite{MaryJane}. When the threshold is lowered, small groups of nodes form and
eventually communities begin to appear. 
In Fig. \ref{figure3}(a) we show the Marvel Universe when the $220$ links with the largest $w_{ij}$ are 
considered~\cite{Pajek}.  Since there are characters with more than one connection, the network is
has only $130$ vertices. 
Four large communities can be clearly distinguished, while the rest of the characters appear connected 
in isolated pairs, or forming very small groups. In these communities two patterns of
interconnections seem to dominate. On one hand some characters form 
tightly knitted groups, as the community on top which includes the character Beast (B) and corresponds to the
X-men. On the other hand, star shaped structures dominated by a central character
can also be clearly distinguished. These central characters are popular characters such as 
Spider-Man (SM) or Captain America (CA). 

\begin{figure}
\includegraphics[width=\linewidth,clip=true]{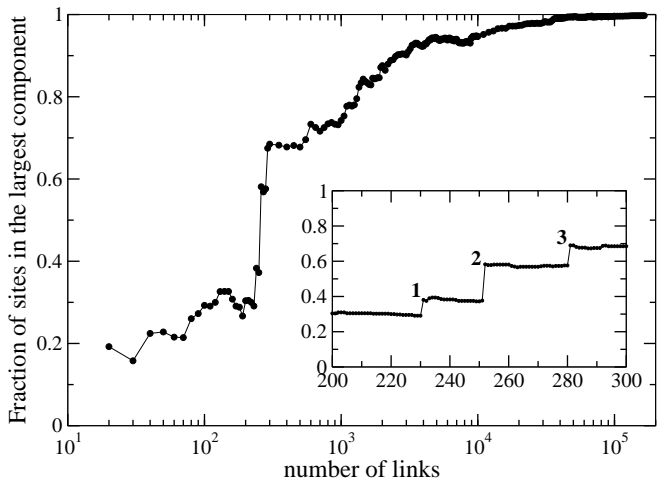}
\caption{\label{figure4} Fraction of sites in the largest component $f_{slc}$ as a function of the number of links added in decreasing 
weight order. The inset shows in detail the transition between $200$ and $300$ links.}
\end{figure}

As the threshold in the weight is lowered further links between communities appear, and
eventually a giant component emerges, as Fig. \ref{figure3}(b) shows when 
$300$ links have been added~\cite{Pajek}.  In order to characterize the growth of the 
network as the
 threshold is lowered, we calculate the fraction of sites in the largest component $f_{slc}$. 
Fig. \ref{figure4} shows the $f_{slc}$ as a function of
 the number of links added in decreasing weight order.  A sharp transition can be observed between 200
 and 300 links, where $f_{slc}$ jumps from less than  $0.3$ to $0.7$.
 After the jump the giant component presents a slow and almost monotonous growth,
 and eventually reaches a saturation value when approximately $40.000$ links are added.

\begin{figure}
\includegraphics[width=\linewidth,clip=true]{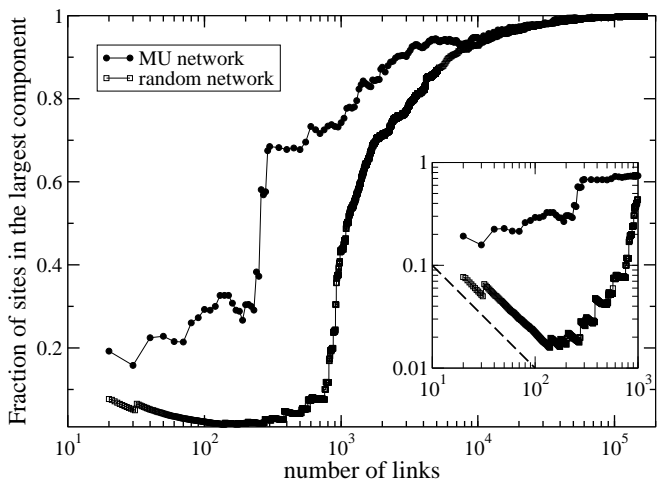}
\caption{\label{figure5} Fraction of sites in the largest component $f_{slc}$
  as a function of the number of links added when the links are chosen in 
decreasing weight order (circles) and also when they are chosen in random order (squares). 
The inset presents a detail for short times in a log-log plot. 
The dashed line shows the power law decay $1/x$ as a guide to the eye.}
\end{figure}

Notice that a very small fraction of links ($ \approx 0.001$) are
necessary for the giant component to emerge. This result suggests a behavior similar 
to a random network. In fact, Callaway {\em et al.} \cite{Callaway}
have shown that if one considers a random network with a truncated power law degree distribution, 
such as the one observed in the MU,  then the percolation threshold is also very small.
However, if one does not take the weights into account and chooses the links in random order, 
thus erasing all correlations, a qualitatively different behavior is
observed. Fig. \ref{figure5} compares the behavior of $f_{slc}$ for the MU and
a typical realization obtained when the links are chosen randomly as a function of the
number of links added. The inset shows the behavior of $f_{slc}$ in a log-log plot. 
Note that when the links are chosen at random  the 
$f_{slc}$ presents regions that decay following a power law close to $1/x$. 
This behavior reveals that new incorporated links do not form part of 
the largest component. They enter connecting isolated pairs of vertices or form part of a
smaller group. When one new link connects a vertex or a group to the largest component, a
jump in $f_{slc}$ is observed. Eventually a giant component emerges and then a
monotonous growth is observed. 

\begin{figure}
\includegraphics[width=\linewidth,clip=true]{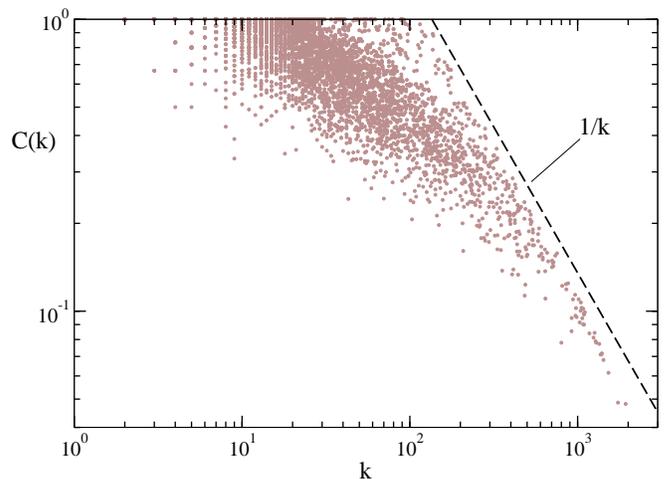}
\caption{\label{figure6} Clustering $C(k)$ as a function of the degree  $k$. The dashed line shows the
  power law $1/k$ as a guide to the eye.}
\end{figure}

The growing behavior observed in Figs. \ref{figure3} and \ref{figure4},  where communities combine  
to form larger but less cohesive structures strongly suggests that the MU network has a hierarchical 
structure \cite{Ravasz_1,Ravasz_2}.  Hierarchical networks integrate both  modular and scale-free structure, 
and can be characterized quantitatively by the scaling law of the clustering coefficient
\begin{equation}
C(k) \sim k^{-1}
\label{eq2}
\end{equation}
where $C$ is the measure proposed by Watts and Strogatz~\cite{Watts} 
\begin{equation}
C = \frac{1}{N} \sum_{i=1}^{N} C_i = \frac{1}{N} \sum_{i=1}^{N} \frac{2n_i}{k_i(k_i-1)}
\end{equation}
Here $N$ is the number of vertices, $k_i$ is the degree of vertex $i$, and $n_i$ is the number of links between 
the $k_i$ neighbors of $i$. 
In Fig.~\ref{figure6} we present the behavior of $C(k)$ as a function of degree $k$ for the MU network. 
The figure  clearly shows the coexistence of a hierarchy of nodes with different 
degrees of clustering. In particular, the nodes with a smaller degree present a higher clustering than 
those with a larger degree, and the decay of $C(k)$ can be bounded by a $1/k$ behavior as the dashed line  shows.

Ravasz and Barab\'asi~\cite{Ravasz_2} note that the presence of a hierarchical 
architecture gives a new interpretation to the role of hubs in complex networks: 
while the nodes with small degree are part of densely interlinked
clusters, the hubs play the important role of bringing together 
the many small communities of clusters into a single, integrated
network. In fact, this seems to be the case in the MU network where, as
we show in the following paragraphs, the most 
popular superheros play the role of connecting the communities.

The sharp jump that marks the appearance of the giant component in the MU (see Fig. \ref{figure4}) shows that the characters
that appear repeatedly in the same publications form a well defined
group. As a consequence, a criterion for setting a cut-off can be well
defined in an analogy with  percolation transitions. If the threshold is set too high 
then the giant component breaks into many small components, such as
the star shaped and the tightly knitted groups. If, on the other hand, the threshold is set too low, 
the slow, almost monotonous growth of the giant component shows that
new characters are  incorporated directly to those already present in
the giant component. As a consequence, as more links are added the
division in communities is harder to determine. Thus, we focus
our analysis of the system close to the transition.

The inset in Fig. \ref{figure4} clearly shows that the transition between $200$ and $300$  links can
be subdivided in three smaller jumps. The events,
 numbered as $1$, $2$ and $3$ correspond to the links between the
 following characters:
\begin{enumerate}
\item Captain America (CA) - Beast (B) 
\item Captain America (CA) - Thing (T) 
\item Hulk (H) - Namor McKenzie (N) 
\end{enumerate}

They are all members of ``The Avengers'', a team of Earth's mightiest heroes, ``...formed to fight the foes no 
single hero could withstand'' \cite{Marvel_teams}.  It is worth
stressing that although these characters form a team, each one clearly belongs to a different
community (see Fig. \ref{figure3}(b)). Note that all the
central characters in each community are linked to other
communities. Also, the central characters tend to be connected between
themselves, forming what is known as a rich-club~\cite{Zhou}. In a rich-club the nodes 
are rich in the sense that they have a large degree. In the MU the rich nodes
also share another property: they are all also ``heroes''. In
fact, as Fig. \ref{figure3}(b) shows, if one labels the characters as heroes or villains~\cite{Heros} one finds
that all the central characters are heroes, while most of the characters that surround them are villains. 
It is also worth stressing that the villains in each community are not connected to villains in other communities.

In his work on the network of collaboration among rappers, Smith notes 
that ``New rap acts often feature prominent names on their most popular 
singles and first albums in order to help attract listeners unfamiliar with them or their style''~\cite{Smith}. 
Perhaps a similar mechanism is present in the MU, where new characters are presented next 
to popular characters so that they may be noticed. This clearly will increase the number
of connections of the most popular characters leading to a rich-get-richer tendency that 
is reflected in their large number of connections. Also since popular characters are also part 
of a team they appear repeatedly together thus forming a rich-club. However, there is another ingredient 
that should also be taken into account, since, as the characterization of the nodes reveals, 
only heroes team up, while villains do not. 

We believe that the origin of this division is due to the fact that, although the Marvel 
Universe incorporates elements from fantasy and science-fiction the
arguments of the stories were restricted by a  set of
rules established in the Comics Authority Code of the Comics Magazine Association of America \cite{Code}.
In particular, rule number five in part A of the code for editorial matter states that 
`` Criminals shall not be presented so as to be rendered glamorous 
or to occupy a position which creates the desire for emulation'', and rule number four
in part B of the same section states that 
``Inclusion of stories dealing with evil shall be used or shall be published only where the intent
is to illustrate a moral issue and in no case shall evil be presented alluringly, nor so as to injure the 
sensibilities of the reader''. As a consequence villains are not destined to play leading roles. 
Also, rule number six in part A states that `` In every instance good shall triumph over evil and
the criminal punished for his misdeeds''. We believe that teams of heroes are formed as a consequence of this
rule. In fact, since the heroes will always eventually win, it is necessary for them 
to show at least that some effort is necessary, and thus they need to collaborate and
cooperate with other superheroes in order to finally defeat their enemies. 

\section{Conclusions}

Summarizing, we analyzed the MU as a collaboration network. 
First we defined the system as a binary network, where a connection
between two characters is either present or absent.  We found that in contrast to 
most real social networks the MU is a disassortative network, with an exponent very similar to the
one observed in a real technological network, the Internet. Then, we used a weight measure to analyze the system 
as a weighted network. This allowed us to distinguish interactions between characters that 
appear repeatedly together to those interactions between characters that meet only few times. 
We observed that the weight distribution presents a power law behavior, and thus a small fraction 
of the interactions are very strong, while the majority interacts very weakly. 
By setting a threshold on the weight we were able to show that the characters that 
appear repeatedly in the same publication form a well defined group. Through the characterization 
of the community structure and also analyzing the clustering as a function of the degree, we showed 
that the network presents a hierarchical structure. 
We also analyzed the role of the hubs, and have shown that these characters form a rich-club of 
heroes that connect different communities. On the other hand characters labeled as villains appear 
around the hubs and do not connect communities. We discussed possible mechanisms that lead 
to these effects.  In particular the rules of the Comic Authority 
Code clearly limit the  role of villains. Also, we believe that heroes need to team up in order 
to show that some effort is necessary to defeat their enemies, since there is a rule that 
states that in the end always good shall triumph over evil.
Finally, we note that a gender classification reveals that all the central characters are males,
 and, as in the case of villains, the female characters do not play a role connecting communities.
 However, as was already noted, the strongest link in the MU is the relation between Spider Man and Mary Jane Watson Parker, 
a fact that shows that although the MU deals mainly with superheroes and villains the 
most popular plot is  a love story.   

\begin{acknowledgments}
The author thanks Sebasti\'an Risau Gusman and Gustavo Sibona for helpful comments and discussions.  This work 
has been supported by grants from CONICET PIP05-5114 (Argentina), ANPCyT PICT03-13893 (Argentina) and ICTP NET-61 (Italy).
\end{acknowledgments}


\begin{thebibliography}{99}

\bibitem{Dorogovtsev} S. N. Dorogovtsev, J. F. F. Mendes, Adv. Phys. {\bf 51}, 1079 (2002).

\bibitem{Albert} R. Albert, and A.-L. Barab\'asi, Rev. Mod. Phys. {\bf 74}, 47 (2002).

\bibitem{Newman} M. E. J. Newman, SIAM Rev. {\bf 45}, 167 (2003)

\bibitem{NBW} M. Newman, A.-L. Barab\'asi, D. J. Watts, {\em The Structure and Dynamics of Networks}, Princeton University Press, 
New Jersey, 2006.

\bibitem{Watts} D. J. Watts, and S. H. Strogatz, Nature {\bf 393}, 440 (1998).

\bibitem{Barabasi} A.-L. Barab\'asi, and R. Albert, Science {\bf 286}, 509 (1999).

\bibitem{Amaral} L. A. N. Amaral, A. Scala, M. Barth\'el\'emy, and H. E. Stanley, Proc. Natl. Acad. Sci. U.S.A. {\bf 97}, 11149 (2000).

\bibitem{Newman2} M. E. J. Newman, Proc. Natl. Acad. Sci. U.S.A. {\bf 98}, 404 (2001).

\bibitem{Newman_pre1} M. E. J. Newman, Phys. Rev. E {\bf 64}, 016131 (2001).

\bibitem{Newman_pre2} M. E. J. Newman, Phys. Rev. E {\bf 64}, 016132 (2001).

\bibitem{Barabasi2} A.-L. Barab\'asi, H. Jeong, Z. Neda, E. Ravasz, A. Schubert, and T. Vicsek, 
Physica A {\bf 311}, 590 (2002).

\bibitem{Girvan} M. Girvan, and M. E. J. Newman, Proc. Natl. Acad. Sci. USA  {\bf 99}, 7821 (2002).

\bibitem{NG} M. E. J. Newman, and M. Girvan, Phys. Rev. E {\bf 69}, 026113 (2004). 

\bibitem{Palla} G. Palla, I. Der\'enyi, I. Farkas, T. Vicsek, Nature {\bf 435}, 814 (2005).		

\bibitem{Danon} L. Danon, A. D\'{\i}az-Guilera, J. Duch, and A. Arenas, J. Stat. Mech.  P09008 (2005).

\bibitem{Gleiser} P. M. Gleiser and L. Danon, Adv. Comp. Sys. {\bf 6}, 565 (2003).

\bibitem{Guimera} R. Guimer\`a, L. Danon, A. D\'{\i}az-Guilera, F. Giralt and A. Arenas, Phys. Rev. E {\bf 68}, 065103(R) (2003).

\bibitem{Arenas} A. Arenas, L. Danon, A. D\'{\i}az-Guilera, P. M. Gleiser, and R. Guimer\`a, 
Eur. Phys. J. B {\bf 38}, 373 (2004).

\bibitem{Boguna} M. Bogu\~n\'a, R. Pastor-Satorras, A. D\'{\i}az-Guilera, and A. Arenas, Phys.  Rev. E {\bf 70}, 056122 (2004).

\bibitem{Motter} A. E. Motter, T. Nishikawa, and Ying-Cheng Lai, Phys. Rev. E {\bf 68}, 036105 (2003).

\bibitem{Noh} J. D. Noh,  H.-C. Jeong, Y.-Y. Ahn, and H. Jeong, Phys. Rev. E {\bf 71}, 036131 (2005).

\bibitem{MarvelUniverse} The Marvel Universe Home Page. http://www.marvel.com/universe/

\bibitem{Alberich} R. Alberich, J. Miro-Julia, and R. Rosell\'o, e-print cond-mat/0202174. 

\bibitem{Chappell} R. Chappell, The Marvel Chronology Project, http://www.chronologyproject.com.

\bibitem{Newman4} M. E. J. Newman, Phys. Rev. Lett. {\bf 89}, 208701 (2002).

\bibitem{Newman5} M. E. J. Newman and J. Park, Phys. Rev. E {\bf 68}, 036122 (2003).

\bibitem{Pastor-Satorras} R. Pastor-Satorras, A. V\'azquez, and A. Vespignani, Phys. Rev. Lett. {\bf 87}, 258701 (2001).

\bibitem{Miro} The data files can be downloaded from http://bioinfo.uib.es/$\sim$joemiro/marvel.html

\bibitem{Li} W. Li, Y. Lin, and Y. Liu, Physica A {\bf 376}, 708 (2007).

\bibitem{MaryJane} For an account on the history of the relation between these characters see
http://www.marvel.com/universe/Watson-Parker\%2C\_Mary\_Jane

\bibitem{Pajek} The figure was drawn using the Pajek Program for Large Network Analysis.
Home page: http://vlado.fmf.uni-lj.si/pub/networks/pajek/

\bibitem{Callaway} D. S. Callaway, M. E. J. Newman, S. H. Strogatz and D. J. Watts, Phys. Rev. Lett. {\bf 85}, 5468 (2000).

\bibitem{Ravasz_1} E. Ravasz, A. L. Somera, D. A. Mongru, Z. N. Oltvai, and A.-L. Barab\'asi, Science {\bf 297}, 1551 (2002).

\bibitem{Ravasz_2} E. Ravasz, and A.-L. Barab\'asi, Phys. Rev. E {\bf 67}, 026112 (2003).

\bibitem{Marvel_teams} The Avengers team is described in http://www.marvel.com/universe/Avengers

\bibitem{Zhou} S. Zhou and R. J. Mondragon, IEEE Commun. Lett. {\bf 8}, 180 (2004).

\bibitem{Heros} The classification is based on information from  http://www.marvel.com/universe/Category:Heroes
and also from http://www.marvel.com/universe/Category:Villains

\bibitem{Smith} R. D. Smith, J. Stat. Mech.  P02006 (2006).

\bibitem{Code} For a short review on the history  and also a link to the complete code see 
http://ublib.buffalo.edu/libraries/projects/comics/cca.html

\end{thebibliography}
\end{document}